\documentclass[letter,11pt]{iopart}

\usepackage{epsfig,amssymb,citesort}

\begin{document}

\title[Magnetic compounds related to the Fe$_2$P structure]{Crystal structure, 
magnetism, and bonding of the hexagonal compounds Pd$_{1.63}$Mn$_{0.37}$Si and 
Pd$_{1.82}$Mn$_{0.18}$Ge related to the Fe$_2$P structure}

\author{Srinivasa Thimmaiah\dag, Claudia Felser\ddag, and Ram Seshadri\dag}

\address{\dag Materials Department and Materials Research Laboratory\\
	 University of California, Santa Barbara CA 93106\\
	 seshadri@mrl.ucsb.edu}

\address{\ddag Institut f\"ur Anorganische Chemie und Analytische Chemie\\ 
Johannes Gutenberg-Universit\"at, Staudinger Weg 9, 55099 Mainz\\ 
felser@uni-mainz.de}

\begin{abstract}
We have used single crystal X-ray diffraction methods to establish the crystal 
structures of a compound in the Pd-Mn-Si system and in the Pd-Mn-Ge system.
The title compounds have structures related to the Fe$_2$P structure type and
are ferromagnetic with Curie temperatures above the room temperature. Density
functional electronic structure calculations help to understand the nature
of the local moment ferromagnetism in these compounds. However neither the 
electronic structure calculations nor the magnetic measurements provide any 
evidence of half-metallic behavior.

\end{abstract}


\maketitle

\section{Introduction}

In 1983, deGroot and coworkers\cite{deGroot} suggested that the electronic 
structure of the ferromagnetic half-Heusler compound NiMnSb possessed a
curious characteristic; at the Fermi level, only states from one spin channel 
were populated; the other spin channel was gapped. They called such compounds 
half-metallic ferromagnets.  The half metallic nature of NiMnSb has found
experimental support from nulk measurement techniques such as 
 positron annihilation.\cite{hansen1,hansen2} However, surface-sensitive 
techniques such as photoemission spectroscopy detect highly reduced spin 
polarization.\cite{zhu,correa} The phenomena of half-metallicity has been 
identified as possessing a history that pre-dates NiMnSb -- for example in 
certain pyrite compounds\cite{CoFeS2_Jarett} and in certain chromium 
chalcospinels\cite{Cr-spinel}, as well as in the famous mixed-valent rare-earth 
manganites.\cite{Jonker} 

Half-metallic ferromagnets are an important materials 
component in the emerging domain of spin-based electronics,\cite{Zutic,Felser} 
and magnetic intermetallics deriving from Heusler and half-Heusler crystal 
structures have been the focus of much recent attention. 
Both the half-Heusler\cite{Whangbo,Rabe,Galanakis_hH,Kandpal} and
Heusler\cite{Kuebler,Galanakis_fH} compounds obey certain simple electron 
counting rules that allow their saturation magnetic moments to be estimated
from the number of valence electrons. One feature that emerges, particularly
for the XYZ half-Heuslers\cite{Whangbo,Kandpal} is that covalency effects in 
parts of the structural network, specifically the YZ zinc blende lattice,
play a crucial role in maintaining the half-metallic gap across a wide range 
of valence electron counts.

It is of interest to explore other equiatomic magnetic compounds with the 
general formula XYZ that are compositionally related to the half-Heusler 
structure but crystallize in other structures. For example, Johnson and
Jeitschko explored a number of silicides and germanides\cite{Jeitschko} 
in the series Mn-Pd-Si and Mn-Pd-Ge. Ba\.{z}ela\cite{Bazela} 
has studied the magnetic properties of some related silicides and germanides 
without establishing the precise composition, assuming that the compounds were 
single phase. Verni\'{e}re \textit{et al.}\cite{Malaman} have reported the 
structure and magnetic properties of PdMnGe by neutron diffraction 
experiments. Eriksson\textit{et al.}\cite{Eriksson} have described a cubic
compound with the formula Mn$_8$Pd$_{15}$Si$_{7}$ in a filled 
Mg$_6$Cu$_{16}$Si$_7$ structure type with complex magnetic behavior, including
non-collinear spin ordering.

We thought it of interest to reexamine compounds in these systems in order 
to determine whether the hexagonal compounds are proximal to half-metallic 
ferromagnetism. Here we report on two compounds which occur as single 
crystals with Fe$_2$P-derived structures in the Pd-Mn-Si and Pd-Mn-Ge phase
diagrams. We have reproduced polycrystalline samples with the compositions 
established by the single crystal structural studies. This has allowed 
magnetic studies on these compounds. In addition, we have performed density 
functional calculations on a typical compound from the Pd-Mn-Si system to 
better understand the nature of the magnetism.

\section{Experimental details and crystal structure studies}

Pd-Mn-X (X: Si and Ge) compounds were prepared from high purity metals
by arc melting in an Ar atmosphere. A composition of 3:1:2
(Pd:Mn:X) was chosen for the preparation. The products obtained 
from arc melting were sealed in evacuated vitreous silica tubes and 
subsequently annealed first at 950$^{\circ}$C for two days and then at 
800$^{\circ}$C for 5 to 10 days in order to improve the site ordering of the 
phases. Crystals were picked from crushed products and subject to single 
crystal diffraction studies. The phase purity of powder samples were checked 
using a Phillips (XPERT MPD, 45\,kV 40\,mA, Cu-K$\alpha$ radiation) 
powder diffractometer, followed by Rietveld refinement of the diffraction
profiles. SQUID magnetization measurements were carried out 
in the temperature range 5 to 400\,K using a Quantum Design MPMS 5XL SQUID 
magnetometer. Zero field cooled (ZFC) and field cooled (FC) magnetization 
as a function of temperature was at a constant field strength of 1000\,Oe. 
In addition, isothermal $M-H$ traces were recorded at different temperatures.

A small crystal of size 0.12 $\times$ 0.07 $\times$ 0.06\,mm$^{3}$ was picked 
from the Pd-Mn-Si sample from a sample that also contained MnSi as a minority 
phase. X-ray diffraction intensities were collected on Bruker CCD diffractometer
(MoK$_{\alpha}$ radiation, $\lambda$ = 0.71073\,\AA) at room temperature. The
measured intensities were corrected for Lorentz and polarization effects and 
were further corrected for absorption using the \textsc{sadabs} 
program.\cite{SADABS} The structure solution was obtained using the 
\textsc{shelxs-97} program and refined using the \textsc{shelxl-97} as 
implemented in \textsc{shelxtl} package using full-matrix least-squares 
refinement.\cite{Smart} The structure was solved in the space group $P\bar62m$ 
(No. 189). The final refinement cycles used anisotropic displacement parameter 
and gave $R_1$ = 0.0264 for 130 unique reflections. There was no evidence
for strong correlation between refined parameters and thermal displacement 
parameters behaved normally. The suggested sample composition was
Pd$_{1.63}$Mn$_{0.37}$Si.

The crystals from the Pd-Mn-Ge system were isolated from the sample consisting 
of Pd$_{3}$MnGe$_{2}$ as a second phase (orthorhombic, $Pnma$, 
$a$ = 6.910(1)\,\AA, $b$ = 3.146(1)\,\AA, 
$c$ = 16.504(4)\,\AA.\cite{Roques,Malaman})
The majority of the sample consisted of the orthorhombic phase. Several 
attempts were made to isolate good quality crystals of the orthorhombic 
phase to confirm the structure independently but these were unsuccessful. 
However, good quality crystals were found to have hexagonal symmetry.
A crystal of size 0.10 $\times$ 0.075 $\times$ 0.06\,mm$^{3}$
was selected for single crystal diffraction studies. The same procedures used
previously were employed to obtain the crystal structure.
The final refinement performed on $F^2$ with 134 unique reflections 
gave R$_{1}$ = 0.0314. The suggested sample composition was
Pd$_{1.82}$Mn$_{0.18}$Ge. 

\section{Computational studies}

An ordered supercell was derived from the structure of Pd$_{1.63}$Mn$_{0.37}$Si
with the composition Pd$_9$Mn$_3$Si$_6$, and with an orthohexagonal
unit cell in the space group $Pmm2$ (No. 25). The calculated composition
can be written Pd$_{1.5}$Mn$_{0.5}$Si and is quite proximal to the 
experimental composition. The electronic structure calculations used
linear muffin tin orbitals (LMTO) within the local spin density approximation, 
as implemented in the \textsc{stuttgart tb-lmto-asa} 
program.\cite{stuttgart1,stuttgart2} 
The calculations were performed on 216 $k$ points within the irreducible 
wedge of the Brillouin zone. The electron localization function 
(ELF)\cite{ELF1,ELF2} has been used to understand the extent to which 
electrons are localized, to locate bonding and non-bonding electron pairs 
in the real space of the crystal structure. 

\section{Results and Discussion} 

\begin{table}
\caption{Crystal structures of the Pd-Mn-Si and Pd-Mn-Ge compounds. 
Space group $P\bar 62m$ (No. 189) obtained from single-crystal X-ray
diffraction.} 
\begin{tabular}{lllllll} 
\hline
\multicolumn{7}{l}{Pd-Mn-Si, $a$ = 6.5033(8)\,\AA\/, $c$ = 3.4622(8)\,\AA\/, 
$R$ = 2.6\%}\\
\hline
Atom   & Wyckoff Symbol & $x$ & $y$ & $z$ & occupancy & $U_{eq.}$(\AA$^2$)\\
\hline
Pd1    & 3$g$ & 0.26315(13) & 0           & $\frac 1 2$ & 1        & 0.013(1)\\
M2(Pd) & 3$f$ & 0.60266(17) & 0           & 0           & 0.633(5) & 0.017(1)\\
M2(Mn) & 3$f$ & 0.60266(17) & 0           & 0           & 0.367(5) & 0.017(1)\\
Si1    & 2$d$ & $\frac 1 3$ & $\frac 2 3$ & $\frac 1 2$ & 1        & 0.016(1)\\
Si2    & 1$a$ & 0           & 0           & 0           & 1        & 0.013(1)\\
\hline
\multicolumn{7}{l}{ \  }\\
\hline
\multicolumn{7}{l}{Pd-Mn-Ge, $a$ = 6.7637(14)\,\AA\/, $c$ = 3.3703(14)\,\AA\/, 
$R$ = 3.2\%}\\
\hline
Atom   & Wyckoff Symbol & $x$ & $y$ & $z$ & occupancy & $U_{eq.}$(\AA$^2$)\\
\hline
Pd1    & 3$g$ & 0.27052(18) & 0           & $\frac 1 2$ & 1        & 0.013(1)\\
M2(Pd) & 3$f$ & 0.6093(2)   & 0           & 0           & 0.823(5) & 0.012(1)\\
M2(Mn) & 3$f$ & 0.6093(2)   & 0           & 0           & 0.177(5) & 0.012(1)\\
Ge1    & 2$d$ & $\frac 1 3$ & $\frac 2 3$ & $\frac 1 2$ & 1        & 0.015(1)\\
Ge2    & 1$a$ & 0           & 0           & 0           & 1        & 0.013(1)\\
\hline
\end{tabular}
\end{table}

\begin{figure}
\centering \epsfig{file=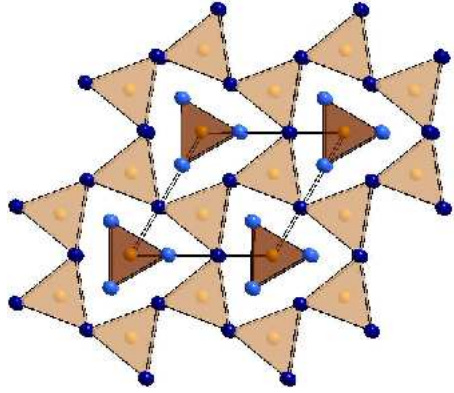, width=6cm}
\caption{Color in online version: Crystal structure of the Pd-Mn-X (X = Si or 
Ge) compounds, showing how X atoms (orange) are in prismatic coordination with
the two metal sites: Pd1 are depicted using light blue spheres and the mixed 
site, M2 is depicted using dark blue spheres. The two kinds of prisms are 
shifted down the $z$-direction by $\frac 1 2$.} \end{figure}

Crystallographic details, fractional atomic coordinates and equivalent 
isotropic displacement parameters are provided for obtained crystal structures
in Table\,1. The crystal structures of PdMnX compounds can be viewed as an 
ordered structure of Fe$_{2}$P with an origin shift of $0 0 \frac12 $, also 
called the ZrNiAl type. The compounds crystallize in the noncentrosymmetric 
hexagonal space group $P\bar 62m$. The compound Pd$_{1.63}$Mn$_{0.37}$Si
is found here to possess the lattice parameters $a$ = 6.5033(8)\,\AA\/ and
$c$ = 3.4622(8)\,\AA, with a $c/a$ ratio of 0.53, nearly the same as the
values reported for Pd$_{1.5}$Mn$_{0.5}$Si by Ba\.{z}ela,\cite{Bazela} 
of  $a$ = 6.4909(6)\,\AA\/ and $c$ = 3.4655(6)\,\AA, also with
a $c/a$ ratio of 0.53.  The lattice parameters determined here for  
Pd$_{1.82}$Mn$_{0.18}$Ge are: $a$ = 6.7637(14)\AA, 
$c$ = 3.3703(14)\AA, with a $c/a$ ratio 0.50. 
However, in the equiatomic PdMnGe reported by Verni\'ere \textit{et al.\/}
the $c/a$ ratio is 0.54.\cite{Malaman} 

The structure of PdMnX is composed of four independent crystallographic sites
(Table\,1).  The position M2($3f$) is a mixed site occupied by Pd and Mn. 
In contrast, the reported equiatomic compound PdMnGe the has the $3f$ position
is completely occupied with Mn. However, there is no PdMnSi compound reported
so far. The projection of the $xy$ plane is shown in figure\,1. 
The structure is built up from X(Si or Ge)--centered tricapped
triangular prisms. The prisms are capped by Pd1 and M2. These triangular
prisms are condensed \textit{via\/} common faces and edges.

\subsection{Pd-Mn-Si}

\begin{figure}\centering \epsfig{file=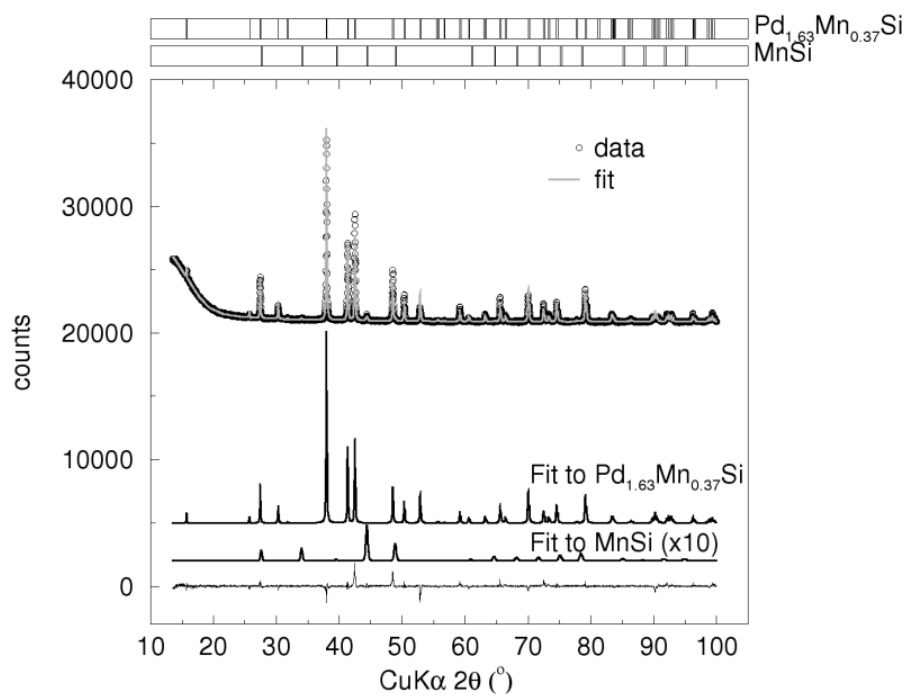, width=12cm}
\caption{Two-phase Rietveld fit of the Pd-Mn-Si phase. 
Points are data, and the gray
line is Rietveld fit. From top to bottom, the traces are data, fit to 
Pd$_{1.63}$Mn$_{0.37}$Si, fit to MnSi (magnified 10 times), and 
the difference profile. Expected peak positions for Pd$_{1.63}$Mn$_{0.37}$Si 
and MnSi are indicated by the vertical bars at the top of the figure.}
\end{figure}

Figure\,2 shows Rietveld refinement of the X-ray diffraction profile of
Pd$_{1.63}$Mn$_{0.37}$Si using the \textsc{xnd} Rietveld code.\cite{xnd}
The starting parameters and site occupancies were taken from single crystal 
data. The lattice parameters after refinement were $a$ = 6.5003(3)\,\AA\/ and 
$c$ = 3.4633(2)\,\AA, in close agreement with the single crystal data. It is 
evident from the fit that the sample contains a noticeable amount of
MnSi.\cite{Rosch} In addition, the difference profile indicates the
possibility of a very small of amount of a third phase Pd$_{2}$Si, which we 
have not accounted for in the refinement. The occurrence of these second phases
arises from the starting elemental ratios of Pd:Mn:Si of 3:1:2 not matching 
with the ratio obtained for the single crystal, which is closer to 5:1:3. 
Johnson and Jeitschko\cite{Jeitschko} have suggested the possibility of 
obtaining single phase samples above 950$^{\circ}$C.\cite{Jeitschko} Our
effort to obtain single phase at 950$^{\circ}$C by quenching the sample
were not fruitful, and once again, the powder diffractogram revealed the 
presence of a second MnSi phase.

\begin{figure}
\centering \epsfig{file=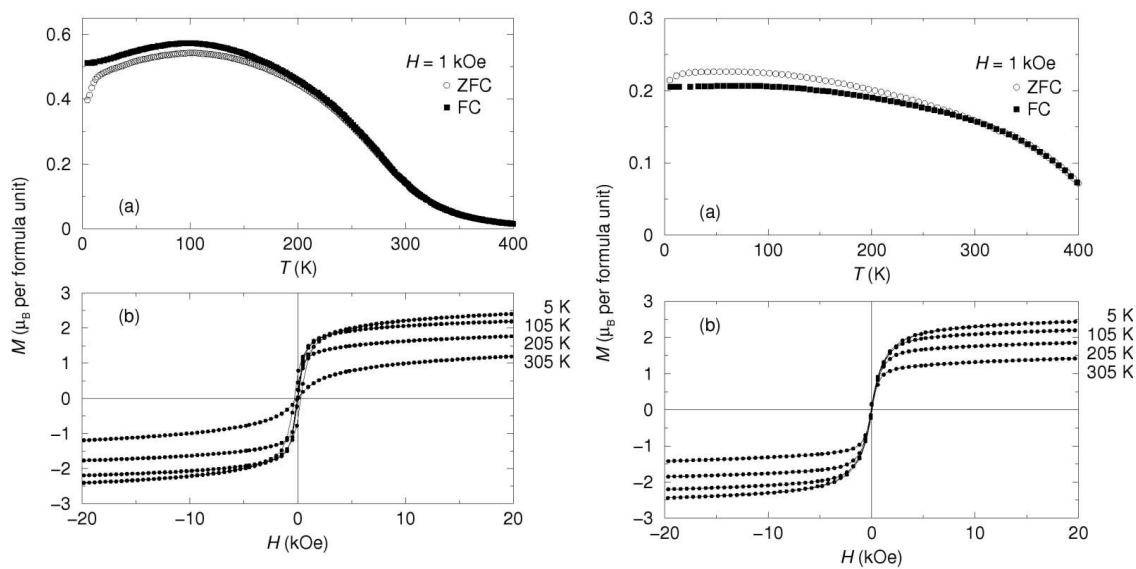, width=15cm}
\caption{(a) Zero-field cooled (circles) and field cooled magnetization 
(filled squares) as a function of temperature, and (b) magnetization as a 
function of field 
at four different (indicated) temperatures. The left panels corresponds
to a sample of Pd$_{1.63}$Mn$_{0.37}$Si prepared by arc-melting of the elements
and then annealed for two days at 950$^\circ$ and 7 days at 800$^\circ$C, 
and the right panels of a sample annealed for 5 days at 800$^\circ$C.} 
\end{figure}

We have studied the magnetic properties of two different samples in the 
Pd-Mn-Si system, both whose X-ray diffraction patterns (Rietveld refined 
lattice parameters) suggest compositions close to that determined by the
single-crystal studies, namely Pd$_{1.63}$Mn$_{0.37}$Si. The samples differ in
the annealing conditions after arc-melting; samples for which data are 
displayed in the left and right panels were annealed as reported in the 
figure caption.  Both samples are soft ferromagnets with Curie temperatures 
above the 400\,K limit of our SQUID magnetometer. Although the detailed
behavior of $M$ \textit{vs.} $T$ for the two samples varies slightly, the 
nature of the $M$ \textit{vs.} $H$ traces shown in panels (b) are quite 
similar, with a 5\,K saturation magnetization near 2.4\,$\mu_B$  per formula
unit. The values reported by Ba\.{z}ela\cite{Bazela}  for the 3:1:2 composition
are $T_c$ = 498 K and the saturation magnetization $M_{sat}$ of 2.12\,$\mu_B$.
This strongly suggests that perhaps the composition in that case as well
is closer to what we have observed in the single-crystal analysis. 
It is important to note that in these samples, the magnetic ion (Mn) is 
relatively diluted and yet the Curie temperature is quite high. Because of
the dilution, it is difficult to associate an integral magnetic moment to 
the composition, and therefore, it is difficult on the basis of the saturation
magnetization alone to determine whether the compound is close to being a 
half-metal. The powder samples have a small, identifiable MnSi impurity.
It is possible that the small downturn in the ZFC trace figure\,3, particularly
in panel (a) is associated with the magnetic transition of MnSi, which is is 
an itinerant ferromagnet with a $T_c$ of 29.5\,K.\cite{Rosch}

\begin{figure}
\centering \epsfig{file=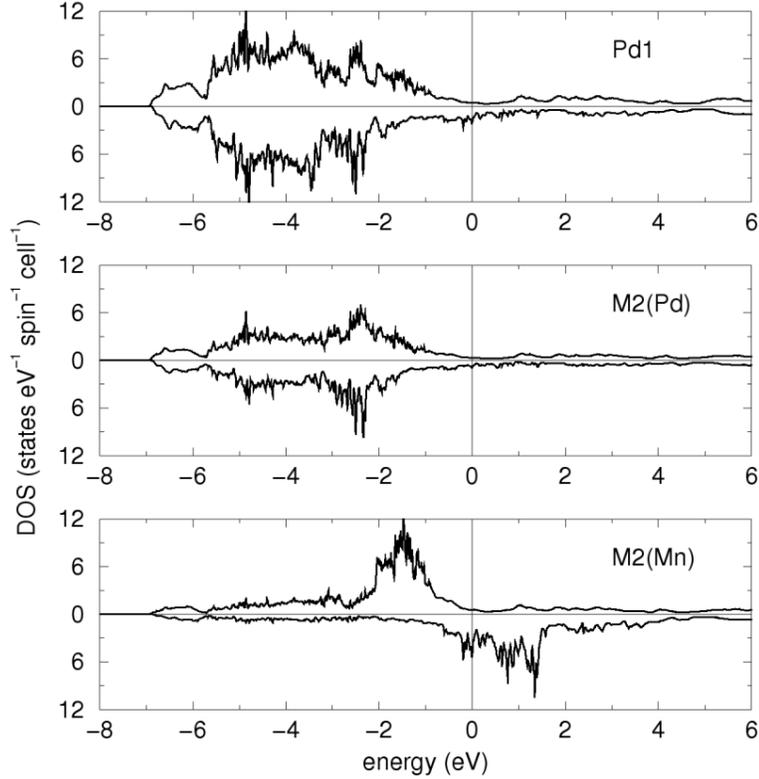, width=10cm}
\caption{Densities of state of Pd$_5$MnSi$_3$ showing in three panels, the
Pd states from the pure Pd1 site, the Pd states from the mixed M2 site, and 
the Mn states, also from the mixed M2 site.} 
\end{figure}

\begin{figure}
\centering \epsfig{file=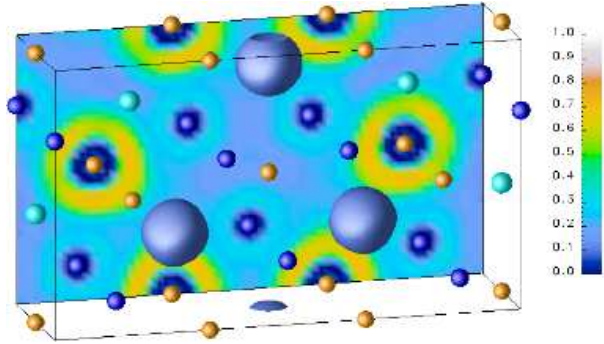,width=8cm}
\caption{Color in online version: Spin density isosurface of Pd$_5$MnSi$_3$ 
for a value of 0.05\,\AA$^{-3}$ showing that the moment is localized on Mn. 
The map at the back of the figure displays the valence-only 
electron localization running from blue (poorly localized) through orange to 
white (strongly localized).} 
\end{figure}

In order to better understand the origin of the magnetism, we have constructed
a orthohexagonal supercell of the Fe$_2$P structure in the space group 
$Pmm2$ (No. 25). LMTO densities of state deriving from the pure and mixed Pd; 
respectively the Pd1 and M2 sites, and from Mn are displayed separately for the 
two spin directions in the panels of figure\,4. It is seen that the system is 
not half metallic since there are both Pd and Mn states in both spin 
directions. Pd seem to have nearly filled d states, as in common for 
intermetallic compounds of Pd with early transition elements.\cite{brewer}
The Mn states are strongly spin polarized suggesting local moment behavior
in this relatively dilute magnetic compound. This local moment behavior is 
better visualized by plotting the spin density in the real space of the 
crystal structure. Figure\,5 displays isosurfaces of spin for a value
of 0.05\,\AA$^{-3}$. It is seen that the spins are well-localized on Mn. In 
addition, figure\,5 shows the valence-only electron localization function 
plotted as a map of values on the rear plane of the unit cell in the level 
where Si 
atoms are found. Their is only somewhat spherical localization around the Si
and no notable localization on any of the internuclear axes suggesting that 
this phase cannot be described using notions of local covalent bonding, unlike
the half-Heusler compounds.\cite{Kandpal} The absence of such covalent 
bonding can perhaps explain the absence of a half-metallic gap in the
title compounds. The calculated magnetic
moment per Pd$_{1.5}$Mn$_{0.5}$Si$_2$ is 1.53\,$\mu_B$, somewhat smaller than 
the experimentally measured value of 2.2\,$\mu_B$ for 
Pd$_{1.63}$Mn$_{0.37}$Si$_2$. The discrepancy could arise because the DFT
calculations are obliged to assume artificial ordering of Mn and Pd. 

\subsection{Pd-Mn-Ge}

\begin{figure}
\centering \epsfig{file=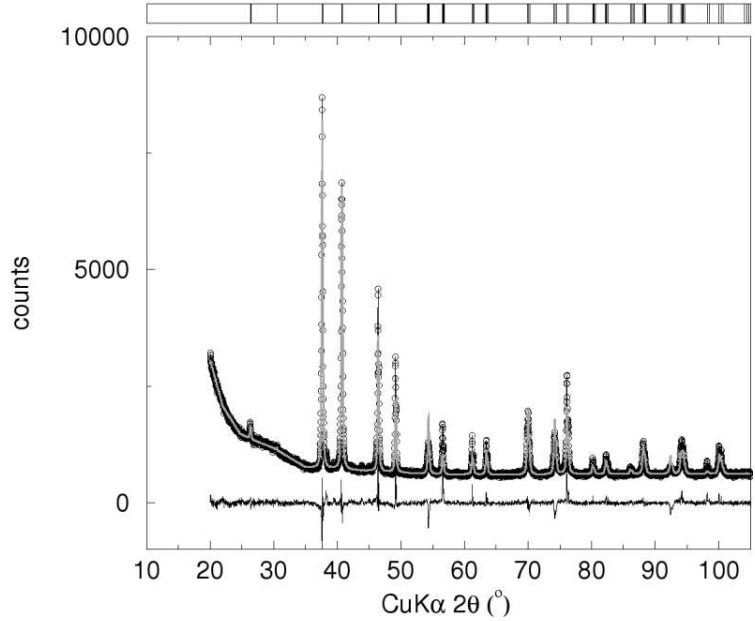, width=10cm}
\caption{Rietveld fit of polycrystalline Pd$_{1.82}$Mn$_{0.18}$Ge. 
Points are data, and the gray line is Rietveld fit. Below these is the 
difference profile. Expected peak positions are indicated by the vertical 
bars at the top of the figure.}
\end{figure}

When polycrystalline samples in the Pd-Mn-Ge systems were prepared with 
starting Pd:Mn:Ge ratios of 3:1:2, the orthorhombic 
Pd$_3$MnGe$_2$\cite{Bazela,Roques} phase was formed, always coexisting with
a second hexagonal phase. When the starting composition was dictated by the
single crystal structural analysis of the hexagonal phase, pure 
polycrystalline samples were obtained that could be subject to Rietveld 
analysis of the X-ray powder diffraction patterns. The lattice parameters 
obtained from the Rietveld analysis were $a$ = 6.7701(6)\,\AA\/ and 
$c$ = 3.3738(4)\,\AA\/ in good agreement with single crystal data. 
Note that the reported lattice constants for equiatomic hexagonal
PdMnGe are $a$ = 6.6453(9)\,\AA\/ and $c$ = 3.36002(5)\,\AA,\cite{Malaman},
suggesting a homogeneity range.

Magnetization as a function of temperature and as a function of field for 
polycrystalline Pd$_{1.82}$Mn$_{0.18}$Ge is shown in figure\,7. 
As a function of temperature under a 1000\,Oe field, it is seen that magnetic
ordering sets in near 360\,K.  However, $M$ \textit{vs.} $H$ traces do not
show clear evidence for magnetic saturation, unlike the related silicide 
compounds, suggesting that while the polycrystalline samples seem phase-pure
by X-ray diffraction, they may not be magnetically homogeneous. At 5\,K and
20\,kOe, the observed $M$ for Pd$_{1.82}$Mn$_{0.18}$Ge is 0.85\,$\mu_B$ 
per formula unit. The values reported for equiatomic PdMnGe ($T_c$ = 541 K
and $M_{sat}$ = 3.5\,$\mu_B$  6\,K)\cite{Malaman} are significantly
different. However, the values we find are not very different from
those determined for Pd$_{3}$MnGe$_2$ ($T_c$ = 392\,K and $M_{sat}$ = 
1.60\,$\mu_B$).\cite{Bazela} If our samples can be considered to be magnetically
homogeneous, then it is possible that the lower $T_c$ and the lower $M_{sat}$
are related simply to the smaller Mn content.

\begin{figure}
\centering \epsfig{file= 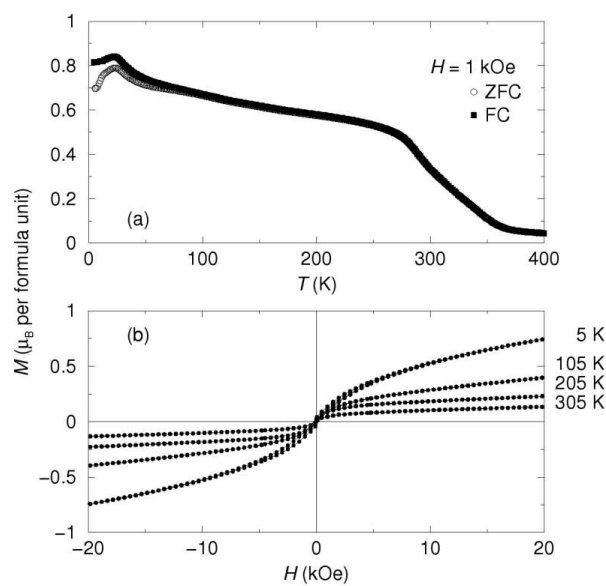,width=8cm}
\caption{(a) Zero-field cooled (circles) and field cooled magnetization 
(filled squares) as a function of temperature for polycrystalline 
Pd$_{1.82}$Mn$_{0.18}$Ge compound.
(b) Magnetization as a function of field for Pd$_{1.82}$Mn$_{0.18}$Ge at
four different (indicated) temperatures.}
\end{figure}

\section{Conclusion} Our interest in half-metallic ferromagnets deriving 
from the equiatomic half-Heusler crystal structure have led us to explore
new compounds in the Pd-Mn-Si and Pd-Mn-Ge systems. The compositions and 
structure of these compounds were established by single crystal methods,
and magnetic studies were performed on polycrystalline samples prepared
according to compositions dictated by the single crystal structure 
determination. The sequence of experiments suggests the importance of single
crystal structure determinations in exploring new magnetic phases. The title
compounds are not half-metallic, perhaps as a result of their not being 
derived from a covalent network with a tendency to semiconducting behavior.
Nevertheless, the silicide compound is interesting because it has a Curie
temperature significantly higher than the room temperature despite having a 
relatively low concentration of magnetic Mn atoms.

\ack{ST and RS gratefully acknowledge the National Science Foundation for 
support through a Career Award (NSF-DMR\,0449354) to RS, and for the use of
MRSEC facilities (Award NSF-DMR\,0520415). We thank Barnaby Dillon for his
contribution, and Dr. Guang Wu for help with single crystal data acquisition.}

\clearpage

\end{document}